\documentclass[letterpaper,twocolumn,10pt]{article}

\usepackage{usenix-2020-09}

\usepackage{tikz}
\usepackage{amsmath}
\usepackage{xspace}

\usepackage[normalem]{ulem}
\usepackage{xurl}
\usepackage{listings}
\usepackage{algorithm}
\usepackage{algpseudocode}
\usepackage{microtype}
\usepackage{balance}
\usepackage{verbatim}
\usepackage{booktabs}
\usepackage{graphicx, subcaption}
\usepackage[]{hyperref}

\newcommand{\shortname}{\textsc{MIGM}\xspace}

\begin{document}

\date{}

\title{\Large \bf Managing Multi Instance GPUs for High Throughput and Energy Savings}

\author{
{\rm Abhijeet Saraha}\\
Georgia Institute of Technology
\and
{\rm Yuanbo Li}\\
Georgia Institute of Technology
\and
{\rm Chris Porter}\\
IBM Research
\and
{\rm Santosh Pande}\\
Georgia Institute of Technology
}

\maketitle

\begin{abstract}
Modern GPUs such as the Ampere series (A30, A100) as well as the Hopper series (H100, H200) offer performance as well as security isolation features. They also support a good amount of concurrency, but taking advantage of it can be quite challenging due to the complex constraints on partitioning the chip.

In this work, we develop partitioning and scheduling schemes for a variety of workloads, ranging from scientific to modern ML workloads, including LLMs. We develop several schemes involving dynamic memory estimation, partition fusion and partition fission. We also support process restart to recover from out-of-memory errors for workloads and early restart as an optimization. This approach yields up to 6.20x throughput and 5.93x energy improvements for general workloads; and we see 1.59x and 1.12x improvement to throughput and energy, respectively, for ML workloads on an A100 GPU. We leverage this technique on LLM workloads and show good improvements, including up to 1.43x throughput improvement and 1.11x energy savings.

\end{abstract}
\section{Introduction}\label{sec-introduction}



Nvidia's \textbf{Multi-Instance GPU (MIG)} feature allows
one to ``slice'' a single GPU into multiple hardware
partitions/segments, yielding
stronger performance isolation and security properties for
multi-tenant workloads.
We explore the problem of \textbf{how to improve
generic workload performance (batch throughput and energy consumption)  by dynamically reconfiguring MIG slices}.
Prior work that leverages MIG has typically looked at performance in terms
of a {\it static} set of slices (e.g. \cite{gpu-beacons-mig}),
or, in the case of dynamic reconfiguration, a more narrow
space that caters to specific problem of managing dynamic partitioning during the inference and retraining phases in continuous learning workloads (attempted in \cite{dynamic-mig-reconfig}).
In contrast, we propose a framework that supports dynamic reconfiguration
for {\it generic} workloads ranging from scientific computing to modern LLM inferencing. Modern workloads involving LLM inferencing have dynamic memory allocations that grow during the workload execution; also the concurrency achieved  is directly influenced by the tightness of the allocated partition to workload's memory needs. This is the key reason why a careful dynamic partitioning scheme must be supported; the complication being MIG partitions are complex to manage (i.e. new partition creation is dependent upon the state of the current configuration of the MIG, Section \ref{sec:partitioning-estimating})

Our  
scheduler has insight into the memory requirements
of each job via \textbf{compile-time analysis, ML model size estimation}
or \textbf{time series-based memory estimation}.
The scheduler receives incoming workloads and places
them on a right-sized
hardware partition which it requests from MIG partition manager.
The partition manager returns a partition that maximizes the flexibility to create future partitions (Section \ref{sec:partition-manager}). Under certain conditions,
the GPU is repartitioned on-the-fly to accommodate the memory and compute needs of the 
workload to be scheduled.

We conducted a preliminary
investigation to understand the importance of
accurate memory predictions and tight partitions to see
the effect on critical metrics such as throughput and
energy consumption.
For this early experiment,
we used the Rodinia benchmark suite \cite{rodinia09}
and an Nvidia A30 GPU. We
drew a random sample of 14 benchmarks to serve as
a batch. We ran the batch twice: once where each
job is assigned to its tightest-fit partition, and a second
where each job is assigned to the next largest
partition.
In this simple experiment,
the throughput (jobs/s) improved 20.6\% and
energy consumption (J) improved 6.3\%, suggesting that
\textit{accurate analysis of jobs' memory footprints
and tight partitions are crucial for performance}.

Summarizing the above discussion, the {\bf main contributions} of this work are as follows:
\begin{enumerate}
    \item A novel time series-based predictive technique for
    determining memory footprints of (practically)
    dynamically unanalyzable jobs, allowing one to still schedule
    these jobs on tight partitions.
    \item Two batch scheduling policies (one that allows reordering of queued jobs and another which maintains the order) with different partition-splitting
    and -merging schemes for addressing compile-time analyzable
    and model size-estimable workloads.  
    \item A dynamic partition manager that manages MIG configurations geared towards maximizing flexibility of future partition creation using a state machine model and its integration with the schedulers.
    \item An integrated system which handles both generic C++ applications
    and PyTorch-based ML workloads
    \item An experimental evaluation showing improvements
    across a range of metrics, including
    memory utilization, throughput, energy, and job turnaround time. 
\end{enumerate}

The remainder of the paper is organized as follows:
Section \ref{sec-background} provides further background and motivation.
Section \ref{sec-memory-prediction} dives into our technique for
estimating the memory usage of machine learning models.
Section \ref{sec-implementation} describes the implementation
details and scheduling algorithms of this system.
Section \ref{sec-evaluation} reports our evaluation results.
Section \ref{sec-related-work} provides additional related work.
Section \ref{sec-conclusion} concludes.

\section{Background \& Motivation}\label{sec-background}



\subsection{GPU Cost, Utilization, and Energy}




High-end GPUs in 2025 cost 3-6x more than their high-end CPU
counterparts. Similarly, renting GPU-enabled virtual machines 
can cost up to 10x more than regular VMs.
GPU utilization is therefore a real
concern, because such costly compute and memory should not be
left idle.
Unfortunately, underutilization continues to be a problem
for GPUs
\cite{patt-util,dl-underutil,keeneland-util,case,gpu-beacons-mig, miso, char-and-pred, mlaas}.
This trend is documented in ML workloads in data centers
\cite{gandiva}, and scientific workloads may see only $\sim30\%$
GPU utilization due to reasons such as varying
kernel sizes \cite{case}.
Furthermore, energy use has become a serious concern today \cite{clover},
both due to
the cost of operating GPUs at warehouse scale and because of
growing concerns over carbon footprint in the community
generally \cite{junkyard}.



\subsection{Tight Partitions and Memory Estimation}\label{sec:partitioning-estimating}

The problem of maximizing utilization and the associated throughput of a MIG device, leading to a reduction in energy consumption, can be defined as generating and allocating the {\it tightest} memory partitions of a MIG device to meet the memory needs of jobs being scheduled. This problem poses several challenges. At the heart of the problem is the need to accurately estimate the peak memory needs of jobs. This must be followed by determining the availability of the tightest partition in the current configuration of the MIG device, or alternatively, creating one if one is not available. Pertinent to the above are scheduling schemes that either leverage the current configuration to the best possible level or perform lightweight dynamic re-partitioning to create the necessary partitions. In case of unknown memory needs, the scheduler starts the process on the smallest memory partition available and in case of an out of memory error, restarts the same on a higher-memory partition. Each of these stages of the solution poses interesting technical challenges which form this paper.

\textit{Memory estimation techniques should be based on
application types.}
Many scientific and image processing workloads can be tackled via \textbf{compiler analysis} ~\cite{case}. Memory needs are determined statically or just in time, before GPU execution begins.
In contrast, ML applications written in popular frameworks
like PyTorch cannot be analyzed via traditional compiler passes.
They are characterized by computation graphs and fixed-size memory pools allocated during training or inference that are dependent on the model, input batch sizes, and a few fixed parameters; such
\textbf{ML model size estimation} can be determined by offline analysis, profiling, and estimation.
This, however, is not sufficient to cover common cases,
including today's LLMs workloads, which allocate memory
dynamically. For such models, offline profiling and memory estimator generation is not possible.
For this we propose a \textbf{time series-based runtime estimator} that projects the process' peak memory needs.

\begin{figure}[ht]
    \centering
    \includegraphics[scale=0.350]{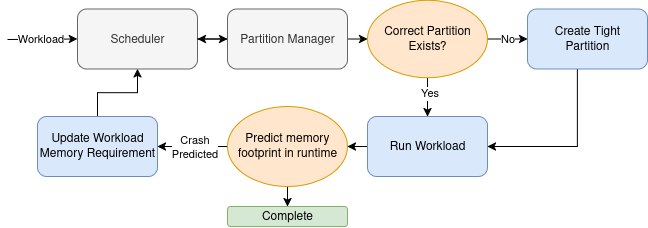}
    \caption{High-level flow diagram of the \shortname framework}
    \label{figure:scheduler-overview}
\end{figure}

\subsection{Overall Approach}

A high-level diagram of our framework, \shortname, is shown
in Figure \ref{figure:scheduler-overview}. The scheduler removes the next workload to be scheduled from the front of a scheduling queue and queries the partition manager to point it to a tight partition.
If a GPU partition is available on current GPU configuration, the workload is launched on the MIG partition. If not, the partition manager will try to dynamically create the MIG partition of the required size. If there are insufficient resources currently available on the GPU (because other workloads are currently running), in case of FIFO scheduling scheme (we call it scheme B), the scheduler will wait until the GPU partition of the correct size becomes available. In order to meet the objective of tightest partition, the scheduler may try to merge or break partitions during this process. 

As mentioned earlier, some workloads whose memory needs are unknown or grow during execution may experience an OOM error at runtime and return to the scheduling queue with updated memory requirements. For such cases, the workload's next run will be on a partition with more memory. Furthermore, to avoid delays due to OOM errors deep into a process' running time, we devise a prediction mechanism which will predict the OOM error and restart the workload on a bigger partition in an early manner. 


\paragraph{Motivating example for ML memory prediction.}
Predicting memory usage in modern ML workloads is 
different than for traditional programs. 
In conventional workloads, memory analysis often focuses
on a single self-contained binary. In contrast,
ML workloads involve a complex interplay between
developer-authored model code, ML frameworks like PyTorch,
and low-level third-party libraries such as cuDNN and cuBLAS. 
This layered architecture obscures memory behaviors,
which are hidden within opaque framework internals
and external libraries, all leading to almost an impossible problem of memory estimation. 

Compiler analysis methods such as those in \cite{case} fall short in practice due to the inherently dynamic
nature of many ML applications. 
For example, large language models (LLMs) used in interactive scenarios dynamically grow their context window as conversations progress, leading to input-dependent tensor sizes and memory allocation patterns that cannot be captured during early-stage profiling.

To address these challenges,  main contribution of this
work is a time series-based memory prediction method. By collecting runtime memory statistics during the initial execution phase,
our system forecasts future peak memory demand.
This allows the scheduler to detect potential OOM risks much earlier, enabling proactive rescheduling of a workload on a larger memory partition. As a result, the system avoids wasting time and GPU resources on partial executions that would ultimately fail.

To test this hypothesis, we experimented
with a Qwen2-7B LLM model.
When it processes increasingly long context windows, the model’s memory usage gradually grows and eventually exceeds the available 10GB of GPU memory after 94 iterations in A100 GPU, leading to a runtime crash due to an OOM error. However, our prediction framework is able to predict that the peak memory usage will surpass 10GB at the 6th iteration. This early warning enables the scheduler to restart the workload on a larger memory partition far in advance of the crash, effectively saving large amount of wasted iterations and ensuring efficient resource utilization.
\section{Memory Usage Prediction for Machine Learning Models}\label{sec-memory-prediction}


In this section,
we begin by analyzing the memory structure of ML models (Section~\ref{sec:mempred-mem-structure}) and explain the extra complexity of ML workload memory structure. Then, we introduce our runtime-driven GPU memory prediction framework (Section~\ref{sec:mempred-dyn-pred}), which combines PyTorch instrumentation and time series forecasting to accurately anticipate peak memory demands in the presence of dynamic memory behaviors.

\subsection{Memory Structure of Machine Learning Workloads} \label{sec:mempred-mem-structure}


Deep learning applications have become the most prominent workloads in modern GPU platforms with high-throughput tensor operations and memory-intensive computations. Accurate memory prediction is therefore critical for allocating tight GPU partitions. 

Existing approaches for GPU memory analysis, especially those developed for traditional high-performance computing applications, are insufficient for machine learning workloads. The fundamental reason lies in the unique memory structure of ML workloads, which is much more layered and complex than that in conventional programs.



An ML workload consists of three components:
\begin{itemize}
    \item \textbf{The model program:} 
    High-level code in ML framework defining the training/inference logic.
    \item \textbf{The ML framework:} The framework handles tensor abstraction, graph construction, scheduling, and dynamic memory management. 
    \item \textbf{Third-party libraries:} These include low-level vendor-optimized libraries like cuDNN, cuBLAS, and CUDA kernels that are responsible for the actual computation 
\end{itemize}

\begin{figure}[ht]
    \centering
    \includegraphics[scale=0.5]{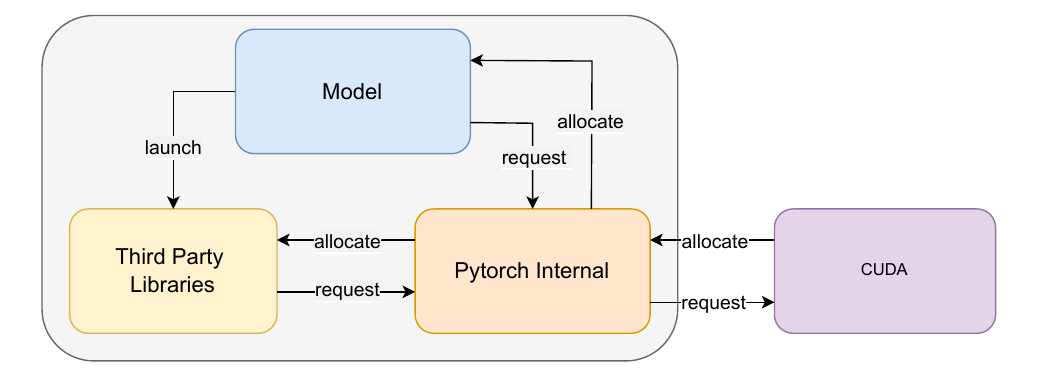}
    \caption{Memory structure for machine learning workloads.}
    \label{figure:ml_memory_structure}
\end{figure}

These three components form a layered architecture where memory interactions are often implicit.
 The model program initiates tensor computations or layer invocations, but it does not interact with CUDA memories directly. Instead it requests GPU memory from the ML framework.
The model may also delegate computations to third party libraries, which may also request workspace from the PyTorch framework to store intermediate tensors.

Figure~\ref{figure:ml_memory_structure} illustrates the memory interaction among these components. ML frameworks such as PyTorch internally maintain a memory manager that handles caching, pre-allocation, and tensor reuse to improve performance. It directly requests memory from CUDA to get GPU memory for the ML workloads. When the model code invokes a layer, it requests memory from the PyTorch framework instead of CUDA directly. Similarly, the model may also launch third-party operators. The PyTorch framework also allocates and re-uses memories. These memory mechanisms may delay or even alter the actual memory allocation behavior, making it difficult to predict the peak GPU memory usage.



\subsection{Dynamic Memory Prediction} \label{sec:mempred-dyn-pred}
To overcome the limitations of static and just-in-time methods when facing opaque and dynamic memory behaviors, we propose a dynamic GPU memory prediction method tailored for PyTorch workloads. 
Instead of analyzing the model alone, our approach captures memory requests within the whole PyTorch framework, enabling early prediction of peak memory size. The core of our method consists of three main components: (1) PyTorch-based instrumentation for tensor memory tracking, (2) workspace memory estimation for third-party libraries, and (3) time series-based forecasting to predict peak usage before it is reached.


\subsubsection{Memory Components in PyTorch Workloads}
Before introducing our tensor tracking approach, we outline the major components of GPU memory in PyTorch workloads:
\begin{itemize} \item \textbf{PyTorch Allocated:} Memory for model tensors (weights, activations, gradients) managed by PyTorch’s caching allocator. This also includes workspace allocated on behalf of \textbf{third-party libraries} (e.g., cuDNN, cuBLAS) used for performance-critical operators.
 
\item \textbf{PyTorch Reserved:} To reduce memory fragmentation and improve performance, PyTorch reserves memory from the CUDA driver in larger blocks and maintains an internal pool. This reserved memory may exceed the amount currently used by tensors, resulting in a gap between physical and logical usage.

\item \textbf{CUDA Context and Miscellaneous:} Overheads from the CUDA runtime and driver. This component is generally fixed for a given GPU and framework version.
\end{itemize}

When executing ML workloads under a memory-partitioned GPU, not all components of total memory usage are equally relevant in determining whether an OOM error will occur. In particular, an OOM error is raised when the memory demand exceeds the physical partition size assigned to the job by the scheduler.

However, it is not necessary for the entire observed memory footprint (as measured by profiling tools like nvidia-smi) to fit within the partition to ensure successful execution. Instead, what truly matters is whether the active memory allocations made by the program exceed the partition limit. This includes:

\begin{itemize}
    \item PyTorch Allocated memory, i.e., memory for tensors (weights, activations, gradients) directly used in computation.
    \item CUDA Context and Miscellaneous memory, which accounts for driver-level and framework initialization overhead.
\end{itemize}

On the other hand, the PyTorch reserved memory is the memory PyTorch pre-allocates and caches for future use, thus not directly causing OOM errors. Therefore, for the purpose of predicting whether an ML workload will trigger OOM errors, what we need to forecast is slightly different than the total memory usage: we must predict the peak PyTorch allocated + CUDA Context/Misc memory during execution.

\subsubsection{Component Memory Usage Estimation}

\paragraph{CUDA context and miscellaneous.}
In practice, this memory consumption is relatively small, and it does not scale with input size or model complexity. It is common to treat the CUDA context memory as a fixed constant per workload. Therefore, the focus of our analysis and prediction lies in capturing the dynamic behaviors of PyTorch allocated memories.

\paragraph{PyTorch allocated memory.}
As previously described in Section~\ref{sec:mempred-mem-structure}, PyTorch’s internal memory allocator is responsible for handling all GPU memory requests.
To accurately capture memory behavior at runtime, we instrument this allocator to record detailed memory usage throughout execution.
This allocator sits at the boundary between the high-level model code and low-level memory APIs, making it an ideal point of intercepting memory behaviors.

In this way, we are able to track all memory requests issued by the model at runtime, including not only memory allocated for user-defined tensors, but also allocations made internally by PyTorch for temporary buffers. 
 This makes our approach significantly more comprehensive than traditional profiling or static analysis techniques, which typically consider only the model-level graph and ignore framework internals or dynamic backend allocations.

\paragraph{Workspace memory estimation.}
To complement our tracking of high-level tensor allocations, we also need to discount the memory size of the third-party workspace.
These workspace sizes usually do not grow with input or context size and thus are excluded from the time series-based prediction.
To estimate this hidden overhead, our framework parses environment variables such as  \texttt{CUBLAS\_WORKSPACE\_CONFIG} to infer the size and count of workspace buffers used by third party libraries.
Our framework walks through model layers, estimates per-layer workspace sizes, and aggregates them to provide a comprehensive view of the total memory reserved for temporary backend use in the workspaces. 

\subsubsection{Time Series-based Prediction on GPU Memory} \label{sec:time-series-est}
\begin{algorithm}[htb]
\caption{Time series-based prediction on peak memory usage.}
\label{alg:predict-peak-mem}
\begin{algorithmic}[]
\Function{PeakMemoryPrediction}{}
    \State $req\_mem\_list \leftarrow$ empty list
    \State $reuse\_ratio\_list \leftarrow$ empty list
    \For{each iteration in ML workload}
        \State collect requested memory $req\_mem$ and $reuse\_ratio$ through instrumented PyTorch.
        \State $req\_mem\_list.\Call{append}{req\_mem}$
        \State $reuse\_ratio\_list.\Call{append}{reuse\_ratio}$
        \State $mem\_mod \leftarrow \Call{fit\_mem\_model}{req\_mem\_list}$
        \State $rt\_mod \leftarrow \Call{fit\_ratio}{reuse\_ratio\_list}$
        \State $mem\_pred \leftarrow$ \\ $ \Call{predict\_peak\_mem}{mem\_mod,rt\_mod, max\_iter} $
        \If{$\Call{converge}{mem\_pred}$}
            \State \textbf{Return} $mem\_pred$
        \EndIf
    \EndFor
\EndFunction
\end{algorithmic}
\end{algorithm}
Given that most ML workloads—especially in training and iterative inference settings—run in a looped, iterative manner, they naturally provide multiple opportunities to observe and analyze their runtime behavior. 
 Specifically, using the memory components described above—PyTorch-allocated tensor memory and estimated workspace memory—we are able to track, at each iteration, the requested memory size observed by PyTorch’s allocator and compute the corresponding reuse ratio, which reflects how effectively PyTorch reuses previously allocated memory blocks.
 With this data, we now describe how we forecast the future peak memory usage of a running ML workload using a time series-based approach.

 \paragraph{Predicting requested memory and memory reuse ratios.}
To forecast the future memory demand of an ML workload, we fit a simple linear regression model to the sequence of observed requested memory values.
Many ML workloads exhibit a gradually increasing behavior in terms of memory, due to accumulating intermediate data, growing context, model states, or cached results.
We use a linear model of the following form to describe the memory usage:

$$\hat{m_t} = a \cdot t + b$$

This is often sufficient to capture the general trend, where $\hat{m_t}$ is the memory request estimated at iteration $t$, and $a, b$ are the learned coefficients. In addition, the linear model is more stable than others when only very few data points are available.

There are also random fluctuations around this potential upward trend due to factors like dynamic memory allocation, batching behavior, or temporary buffers. 
To capture this variability, which is essential to predicting the peak memory usage, it is crucial to model not only the trend but also the stochastic nature of these fluctuations.
The key to model the fluctuations is to analyze the residuals, representing the differences between the actual observed memory and the predicted values from the linear model. By assuming a normal distribution on the residuals, we are able to construct a 99\% confidence interval (CI) for future memory predictions, effectively accounting for both the trend and the variability of the observed data.

The final predicted peak memory request at a future iteration $t$ is:

$$mem\_pred=a\cdot t + b + z\cdot \sigma$$

The term $z$ is the $z$-score corresponding to the desired confidence level, and $\sigma$ represents the standard deviation of the residuals.

In addition to forecasting the requested memory, we also model the memory reuse ratio, which reflects how efficiently memory is reused during execution.
A lower reuse ratio indicates more reuse, meaning that the actual physical memory needed is smaller relative to the total requested memory. 
Empirically, this ratio tends to decrease over time as the workload grows more tensors can be freed and reused.
To fit this behavior using the same linear modeling framework, we transform the reuse ratio by taking its reciprocal, referred here as the inverse reuse ratio, i.e. $inv\_reuse = 1 / reuse\_ratio$.
Using the same approach as requested memory estimation, we can fit the linear model to the inverse reuse ratio, thus predicting future memory reuse efficiency and infer the expected physical memory demand more accurately.

\paragraph{Overall prediction algorithm.}
Algorithm~\ref{alg:predict-peak-mem} presents our overall time series-based algorithm to predict the peak memory usage.
The algorithm begins by initializing two empty lists: one for recording the requested memory at each iteration ($req\_mem\_list$) and another for the memory reuse ratio ($reuse\_ratio\_list$).
For each iteration during machine learning tasks, we collect the current requested memory and reuse ratio through our instrumented PyTorch runtime and append them to their respective lists.
 Using the collected memory data, we fit a linear regression model for requested memory and reuse ratio respectively as described above.
 We then combine the two models to predict the peak memory usage for the final iteration.
 After each prediction, we check for convergence for our prediction. When a convergence is detected, the memory estimator reports the predicted peak memory usage.
\section{Scheduler and Partition Manager}\label{sec-implementation}
Before we discuss the partition manager and scheduler, we introduce the architecture of a typical MIG. For this work, we have chosen A100 GPU which is the state of the art MIG used in industry. 
\begin{figure}[ht]
    \centering
    \includegraphics[scale=0.22]{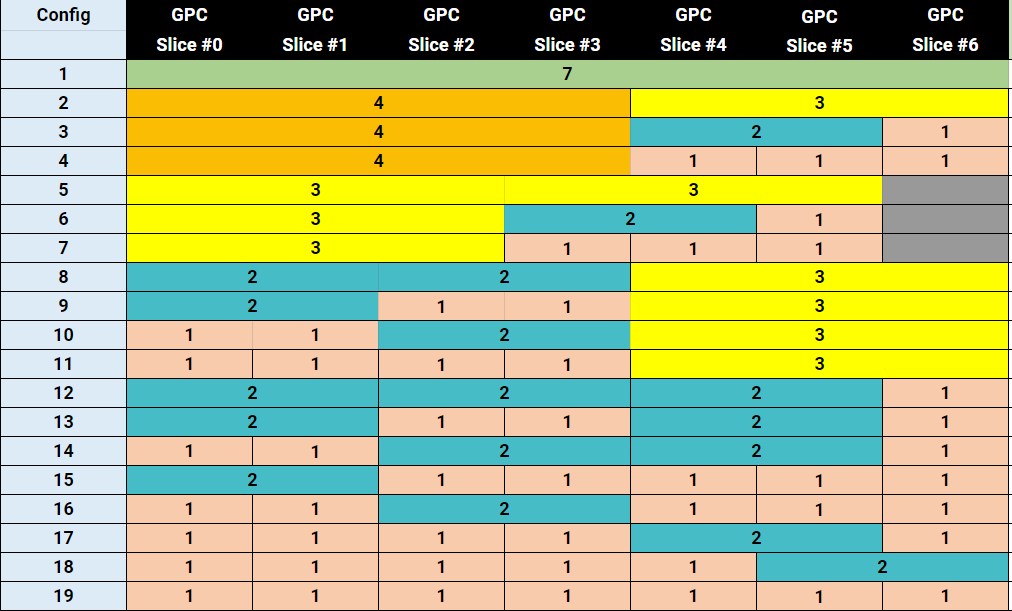}
    \caption{A100 Configurations}
    \label{fig:a100-configs}
\end{figure}
\vspace{-2.5em}
\subsection{A100 Architecture }

Figure ~\ref{fig:a100-configs} shows different configurations which are possible on NVIDIA GPU A100. 
 On this GPU, the hardware supports only a fixed set of valid partition configurations. Each partition will correspond to a particular combination of computation resources and memory slices. For example, the A100 40GB GPU can be partitioned into the following sizes: (1) 1/7 of compute, 5 GB memory; (2) 2/7 of compute, 10 GB memory; (3) 3/7 of compute, 20 GB memory; (4) 4/7 of compute, 20 GB memory; and (5) the full GPU with all compute and memory.

While these profiles provide flexibility in serving workloads of different sizes, they can only be combined into a limited number of valid full-GPU configurations predefined by the hardware. For example, when the MIG is configured with (5GB, 5GB, 30GB unallocated) memory parition, it can only allocate a 20GB memory partition as (5GB, 5GB, 10GB unallocated, 20GB), and it is illegal to have the partition of (5GB, 5GB, 20GB, 10GB unallocated). 



\subsection{Partition Manager}\label{sec:partition-manager}
MIG-enabled GPUs support several specific partition layouts.
A partition is valid only if there is a valid configuration 
it can be extended to. For example, in the 40GB Nvidia A100 GPU,
a partition state of (5GB, 5GB, 30GB-unallocated)
is a valid partition state, because it can be extended to valid
configuration, e.g. (5GB, 5GB, 10GB, 20GB) \cite{mig-user-guide}.


Given a partition state of the GPU and a new partition size request, the placement of the new partition can be interpreted as a state transition in a finite-state machine (FSM). 
Each state corresponds to a valid partition state, and each transition represents the allocation (or deallocation) of a partition.
Since there may be multiple valid ways to serve a new request, i.e. placing a new partition in different GPU slices, one must devise a strategy that allows maximum flexibility to create and allocate partitions described below.

\paragraph{Partition allocation/deallocation algorithm.}

\begin{algorithm}[htb]
\caption{Precompute future-configuration reachability for MIG partition states}
\label{alg:precompute-reach}
\begin{algorithmic}[]
\Function{precompute\_reachability}{}
  \State Enumerate all valid partition states $S$.
  \State Initialize the future configuration reachability mapping $fcr$, 
  \For{each valid partition state $s$}
    \State Compute all reachable fully configured states $F_S$
    \State $fcr(s) \gets |F_s|$.
  \EndFor
  \State
  \Return $fcr$
\EndFunction
\end{algorithmic}
\end{algorithm}

\begin{algorithm}[htb]
\caption{Online allocation by maximizing future reachability}
\label{alg:online-alloc}
\begin{algorithmic}[]
\Function{allocate\_partition}{$s, x, fcr$}
  \State $C \gets \Call{enumerate\_placements}{s, x}$
  \If{$C = \emptyset$}
    \State \Return \textsc{FAIL}
  \EndIf
  \State $s^\star \gets \Call{argmax}{t \in C,\ fcr[t]}$
  \State \Return $s^\star$
\EndFunction
\end{algorithmic}
\end{algorithm}

To maximize hardware parallelism and resource utilization, partition allocation is guided by the transition that preserves the greatest flexibility for future allocations. We quantify this flexibility with the future configuration reachability metric, defined as the number of valid fully configured MIG states that remain reachable from the current state through legal partition allocations. Since the number of states is finite, future configuration reachability can be precomputed offline for all valid states (Algorithm~\ref{alg:precompute-reach}). Online de-allocation is trivial, thus we only discuss the allocation algorithm. During online allocation (Algorithm~\ref{alg:online-alloc}), when multiple placements are feasible, we select the successor state with the highest future configuration reachability value. This ensures that each allocation keeps as many future configuration options open as possible, thereby maintaining high adaptability to diverse workloads.

\paragraph{Formal definition of the Partition State Machine.}

We define the \emph{Partition State Machine} as an FSM:  
\[
\mathcal{M} = (S, \Sigma, \delta, s_0, F)
\]
where:

\begin{itemize}
  \item \( S \) is a finite set of valid partition states of the GPU, e.g. \((5\text{GB}, 5\text{GB}, 30\text{GB-unallocated})\) in an A100 GPU.
  \item \( \Sigma \) is the finite input alphabet. Each input represents a partition allocation or deallocation action. In our case,  
  \[
  \Sigma = \{\texttt{alloc}(x), \texttt{free}(x) \mid x \in \mathcal{P} \}
  \]  
  where \( \mathcal{P} \) is the set of all valid MIG partition sizes, e.g. 5GB, 10GB, and 20GB in an A100.
  \item \( \delta: S \times \Sigma \rightarrow S \) is the transition function. Given a current state and an allocation/deallocation action, it returns the resulting state if the action is legal, or is undefined otherwise.  
  \item \( s_0 \in S \) is the initial state, typically the unpartitioned GPU, e.g. \((40\text{GB-unallocated})\) for an A100.  
  \item \( F \subseteq S \) is the set of final (fully configured) states, corresponding to complete MIG configurations, e.g. \((10\text{GB}, 10\text{GB}, 20\text{GB})\).
\end{itemize}

\paragraph{A100 example of partition allocation.}

Consider a 40GB Nvidia A100 GPU where the current partition state is 
(40GB-unallocated), and a new request for a 5GB partition arrives. There are multiple valid ways to satisfy this request, each leading to a different next configuration \cite{mig-user-guide}:

\begin{itemize}
  \item \((5\,\text{GB},\ 35\,\text{GB-unallocated})\): allocate to the first slice. 
  \item \((5\,\text{GB-unallocated},\ 5\,\text{GB},\ 30\,\text{GB-unallocated})\): allocate to the second slice. 
  \item ...
  \item \((35\,\text{GB-unallocated},\ 5\,\text{GB})\): allocate to the last slice.
\end{itemize}


While all options are valid, they differ in their \emph{future configuration reachability}—the number of legal configurations that can be reached from each state by further partitioning. Specifically, their reachability scores are:

\begin{itemize}
  \item \((5\,\text{GB},\ 35\,\text{GB-unallocated})\): 7 reachable configuration.
  \item \((5\,\text{GB-unallocated},\ 5\,\text{GB},\ 30\,\text{GB-unallocated})\): 7 reachable configurations.
  \item ...
  \item \((35\,\text{GB-unallocated},\ 5\,\text{GB})\): 9 reachable configuration
\end{itemize}

Allocating to the last slice has the largest future configuration reachability, thus offering greatest flexibility for future partition requests. In contrast, the other two configurations lead to fewer final configurations, thus less flexible.

By selecting the transition with the \textbf{highest} \emph{future configuration reachability} (in this case placing the new partition on the second slice), we maximize the number of future options and preserve higher potential for parallel execution. 
This example illustrates how our transition scheme helps avoid premature resource fragmentation and promotes sustained high utilization.

\subsection{Scheduler and Scheduling Algorithms}

\paragraph{Resource estimation for scheduling}
Our scheduler leverages \textbf{compiler analysis} through \cite{case} to get the memory and compute resource requirement (warps) of general scientific workloads, such as  Rodinia, during runtime. We choose the MIG size for
each job such that all warps and the max memory footprint could be
supported.
Although compute is a soft constraint, taking it into account while determining the tightest fit MIG size for the workload prevents degradation of individual benchmark runtime. In orderto maximize concurrency, we perform warp folding as an optimization. For example, consider a workload that needs more SMs (streaming multiprocessors) (say 120) than a GPU can provide (say 100). The workload will execute for 2 time steps assuming ideal parallelism. By allocating only 60 SMs to the workload, one is still able to maintain 2 timestep completion time of the workload but free 40 SMs to allocate other workload. Such optimization indeed allows fitting a workload to the available configuration of GPU. 

For deep neural network benchmarks, we leverage the DNNMem framework \cite{dnnmem} for offline model size estimation as the starting size of the MIG slice for a workload. In case there is an OOM error, the scheduler handles it by rescheduling the workload on the next largest slice. For example, if a workload running on a 10GB slice experiences an OOM error, the framework reschedules the same on a 20GB memory slice. 

For machine learning models that show dynamic memory usage, however, we use time series estimation as described in Section \ref{sec-memory-prediction}. In case the predicted requirement of memory goes over the size of the allocated MIG slice, the workload is preempted and rescheduled on the slice that meets the memory requirement. 
\paragraph{Scheme A: Scheduling by size}
The key goal of this scheme is to minimize the number of dynamic reconfigurations. The algorithm first analyzes the workload queue and sorts it in the order of increasing memory demands. Next, it forms slices corresponding to the smallest memory workloads by invoking the partition manager. Then it schedules the jobs concurrently. That is, the scheduler creates seven 5gb partitions and schedules all small jobs (<5gb slice).

It keeps scheduling on this configuration until all workloads with current memory requirement finish executing. At this point, it reconfigures the GPU with next larger partitions, as per the state transitions embraced by the partition manager, and keeps repeating the above process until all jobs are scheduled. In this manner, it minimizes reconfigurations of the GPU. The pertinent scheduling algorithm is shown in Algorithm \ref{alg:policy-a}; the reconfiguration calls are handled in the background by the partition manager when a slice request of a given size cannot be fulfilled under the current partition. Since the partition sizes in a given configuration are of the same size, the scheduling of jobs (of same size) is multi-threaded and lock free for efficiency reasons.

\paragraph{Scheme B: Scheduling in order}
Algorithm B schedules jobs in order of their arrival in the job queue, in order to maintain fairness. Appropriate GPU partitions are created as per the requirement of the current job being processed by the scheduler. 

The partition manager maintains an updated view of the MIG partitions on the GPU. The scheduler uses the current state of partitions to find an idle partition that tightly fits the current job. If such a partition is unavailable, the scheduler then tries to create a new partition as per the resource requirement of the job in consideration. If the creation of partition fails, then the scheduler uses the partition manager to \textit{merge} neighboring small partitions or \textit{split} bigger partitions to create the tightest fit partition for the current job. If there are no partitions to merge/split then the scheduler waits for a job currently running on the GPU to finish, before trying to find or create a new partition.


\begin{algorithm}[htb]
\caption{Pseudocode for scheme A's scheduling in groups based on MIG slice sizes.}
\label{alg:policy-a}
\begin{algorithmic}[]
\Function{schedule\_by\_group}{$workload$}
  \State $wl\_groups \gets \Call{sorted\_by\_mig\_group}{workload}$
  \For{$group$ in $wl\_groups$}
    \State $\Call{set\_homogeneous\_slices}{group}$
    \State $\Call{schedule}{group}$
  \EndFor
\EndFunction
\end{algorithmic}
\end{algorithm}

\begin{algorithm}[htb]
\caption{Scheme B pseudocode for dynamic reconfiguration scheduling.}
\label{alg:policy-b}
\begin{algorithmic}[]
\Function{schedule\_dyn\_reconfig}{$workload$}
  \While{$workload$}
    \State $j \gets workload.\Call{pop}{ }$
    \While{$true$}
      \State $success \gets \Call{try\_schedule}{j}$
      \If{$success$}
        \State \textbf{break}
      \EndIf
      \State $success \gets \Call{try\_new\_mig\_slice}{j.memfp}$
      \If{!$success$}
        \State $\Call{sleep}{ }$
      \EndIf
    \EndWhile
  \EndWhile    
\EndFunction
\end{algorithmic}
\end{algorithm}

\section{Evaluation}\label{sec-evaluation}

Our testbed consists of an A100 40GB PCIe GPU served
by dual-socket Intel Xeon Platinum 8352Y 32-core processors
with 256GB RAM on the Rogues Gallery testbed \cite{young:2019:rg-exp-insights}. We use \textit{nvidia-smi} command line utility to poll the GPU for power draw (needed for energy calculation) and memory usage every 0.1 seconds (fastest polling rate).  
Our benchmarks consist of Rodinia v3.1 \cite{rodinia09,rodinia10},
the set of ML workloads from \cite{dnnmem}
(with PyTorch v2.8.0),
and 4 LLM workloads: FLAN-T5 training \cite{flan-t5},
and inference on FLAN-T5,
Qwen2-7B \cite{qwen2}, and Llama 3-3B \cite{llama3}.
The baseline scheduler for all
experiments is a non-partitioned A100 GPU
that executes a single workload at a time from the batch ie, the batch executing sequentially on the GPU. 

Our mixes are further detailed in \ref{sec:workload-details}.
We run 7 different Rodinia mixes, selected from a population
of 23 benchmark+parameter combinations, which fall into
4 bucket sizes for the A100: small, medium, large, and full.
These correspond to the 5GB, 10GB, 20GB, and 40GB partition
sizes. We express these mixes in terms of ratios in the
evaluation using the form ``small:medium:large:full'', e.g.
a 4:0:1:1 mix. 
We also run 3 mixes of ML workloads from \cite{dnnmem}, which
consist of several deep neural network benchmarks:
vgg16, resnet50, inceptionv3, and bert.
Lastly, we run homogeneous mixes for all of the
LLM workloads for exercising time series-based prediction.

Our evaluation aims to answer the following questions:
\begin{enumerate}
    \item How does \shortname perform on different
    types of workloads, including those that dynamically
    allocate memory?
    \item How do \shortname's scheduling policies
    compare against a baseline scheduler and each other
    for key metric such as throughput and energy usage?
    \item How accurate is the time series-based predictor,
    and how much does it improve over a non-predictive mechanism?
\end{enumerate}

\subsection{General Workloads}


To exercise the compiler-based analysis alongside the
scheduling component of \shortname, we run several mixes of Rodinia.
Figures \ref{figure:throughput-rodinia}-\ref{figure:turnaround-rodinia}
depict 4 critical performance
metrics: throughput (jobs/sec), energy consumption (J), memory
utilization (\% of GPU memory), and job turnaround time (s), normalized against the baseline.

The throughput improvement is shown in
Figure \ref{figure:throughput-rodinia}.
The homogeneous mixes (Hm1-4) perform better
on the whole.
Hm4 is a mix of only euler3D jobs, which occupies the 20GB slice
(i.e. half of the A100). For this reason, its maximum possible
throughput improvement is 2x, and achieving  $\sim$ 1.7x for both scheduling policies is promising.
Hm2 and Hm3 are gaussian and myocyte mixes, respectively. These
occupy the 5GB slices, and the A100 can support up to 7 simultaneous
jobs; despite some resource contention, these
mixes receive substantial benefits (up to 6.2x).

We break down and compare time spent 
in different stages of the run for the same
workload in Hm3 (myocyte) for both baseline and scheme A (further detailed in Table \ref{tab:myocyte-breakdown} in \ref{sec:workload-details}).
Metrics like GPU kernel runtime remain comparable between the two runs, but there is a noticeable increase in time spent during GPU memory de/allocation,
which negatively impacts throughput. Hm1 runs workloads with 7 MIG instances concurrently in scheme A. As MIG provides full physical isolation through the entire memory system \cite{mig-user-guide}, the extra bookkeeping for each slice during memory management incurs overhead.   

As observed in \cite{pcie},
PCIe bandwidth remains a shared resource, being equally divided
among multiple MIG instances. 
This can cause contention when running multiple workloads that require high PCIe bandwidth to transfer data between the host and device.
To test this, we run a homogeneous mix (batch size 21) of Needleman-Wunsch workloads with initial arguments such that each workload fits into the smallest MIG slice on A100. We see a 1.92x improvement in throughput, as opposed to the theoretical max of 7x. This is explained by the \textasciitilde2.2x increase in runtime of each individual workload running with scheme A vs. baseline.
Profiling the workload, we observe that it spends a significant part of total runtime in data transfer to and from the host. The improvement in throughput occurs by parallelizing part of time the benchmark spends executing GPU kernel.

The heterogeneous mixes (Ht1-3) are formed by taking
different benchmarks and parameter combinations from
the Rodinia suite and randomizing the order of the mix.
Ht2 contains an equal number of small, large and full jobs.
It shows an improvement of 4\% for scheme B and 14\% for scheme A.
Ht3 contains the same number of medium and full jobs, while increasing the number of small jobs by 3x. The improvement in throughput increases to 21\% in scheme B and 29\% in scheme A. This shows that increasing the number of small jobs increases the opportunity for concurrency. Lastly, Ht1 is a mix of small, medium and full jobs such that the total execution time of all 3 groups is roughly the same. We see an improvement of 47\% in scheme B and 64\% in scheme A. scheme A consistently performs better for heterogeneous batches of workloads because scheme B schedules the workloads in order to maintain fairness. For example, if a workload that occupies half the GPU is running and the next job requires the full GPU, scheme B would wait for the first workload to finish, even though there might be workloads that can fit on the idle half of the GPU in the queue. This incurs loss in possible concurrency, depending on the order of incoming workloads.

Energy savings, memory utilization, and job turnaround time
follow the trends in throughput; we make a few key points.
The first is that job turnaround time is significantly better for
the heterogeneous mixes for scheme A; this is because small jobs
(which also take less time to run, generally) are always executed
first. Another point is that the energy savings tracks closely
with the throughput improvements.
Memory utilization is better across all mixes, especially for
homogeneous mixes. Finally, scheme A performs better in general.
This is mostly due to the fact that it is unfair (within a batch),
and thus utilizes its partitions effectively before changing to
another layout.

\subsection{ML Workloads}

\subsubsection{Deep Neural Net Workloads}

We train VGG16, ResNet50, InceptionV3, and BERT
using the same datasets as \cite{dnnmem}.
For these non-LLM jobs, \shortname relies on model size estimation
 techniques in DNNMem \cite{dnnmem}. 
By DNNMem framework estimation, VGG16, ResNet50 and InceptionV3 occupy the 20GB MIG slice, while BERT can occupy either a 5GB or 20GB slice with different batch size and sequence length. We test 3 different jobs mixes: Ml1 contains equal number of small and large jobs, Ml2 contains only small jobs and Ml3 contains only large jobs. (See also Table \ref{tab:ml-mixes} in \ref{sec:workload-details} for mix details).

As shown in Figure~\ref{figure:throughput-ml}-~\ref{figure:turnaround-ml}, all mixes show improvement in metrics for scheme A or scheme B. There is 58\% improvement in throughput, and 12\% improvement in energy consumption for Ml2 running on scheme A, and 43\% improvement in throughput and 5\% in energy consumption while running on scheme B. The improvement in throughput is not close to the theoretical ceiling of 7x. As discussed in 5.1, workloads with high data transfer between host and device experience degradation in runtime if put on a smaller MIG slice, even if the MIG slice satisfies the memory and compute requirement of the workload. Since training deep neural networks is highly data transfer intensive, we observe a less than optimal improvement in throughput in Ml2 and Ml3. Longer and similar runtimes of models in Ml2 that almost saturate the 5gb MIG instance (\textasciitilde3.5gb and \textasciitilde4.7gb) is responsible for high improvement in memory utilization. 
 
For Ml3, throughput improvement is 24\% over baseline for scheme A and 43\% for scheme B. This is the only corner case where scheme B performs better than scheme A. From section 4.1 we have observed that when A100 is partitioned into two MIG instances of 20gb each, the first MIG instance gets 4/7 of the compute resources and the second MIG instances get 3/7 of the compute resources. The multi-threaded implementation of scheme A, as described in 4.3, equally divides the number of jobs to be scheduled on the two partitions. The thread scheduling on the first half of the GPU completes its half of the jobs faster, leading to this corner case of slight loss in concurrency and throughput.

\subsubsection{Dynamic Memory Prediction}

Across the dynamic workloads, we observe that the use of memory predictions provides consistent improvements over both the baseline and policies without predictions. We discuss these improvements metric by metric.

\begin{figure*}[ht]
  \begin{subfigure}[b]{0.3\textwidth}
  \centering
      \includegraphics[width=\textwidth]{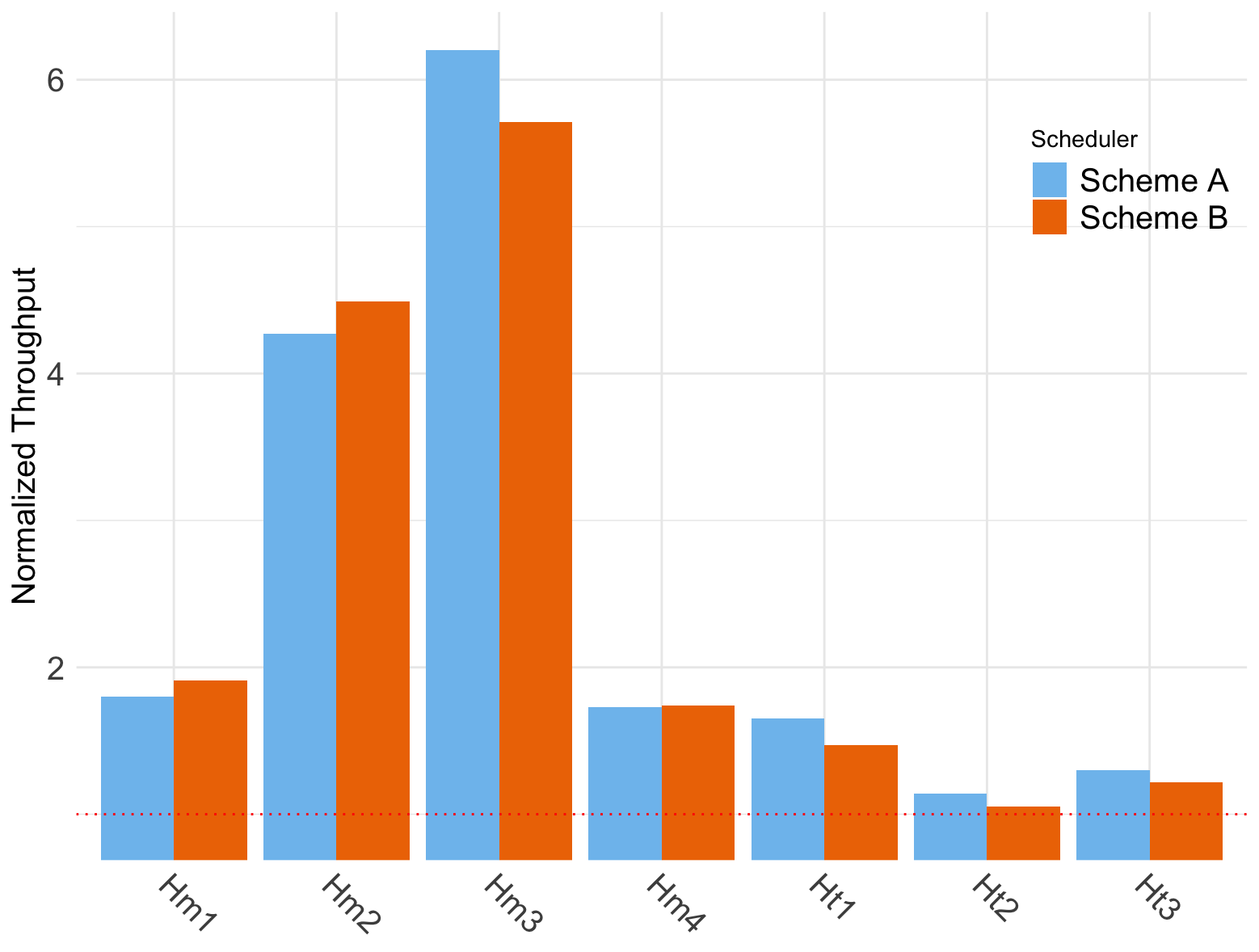}
    \caption{Throughput - Rodinia}
    \label{figure:throughput-rodinia}
  \end{subfigure}
  \qquad
  \begin{subfigure}[b]{0.3\textwidth}
      \centering
      \includegraphics[width=\textwidth]{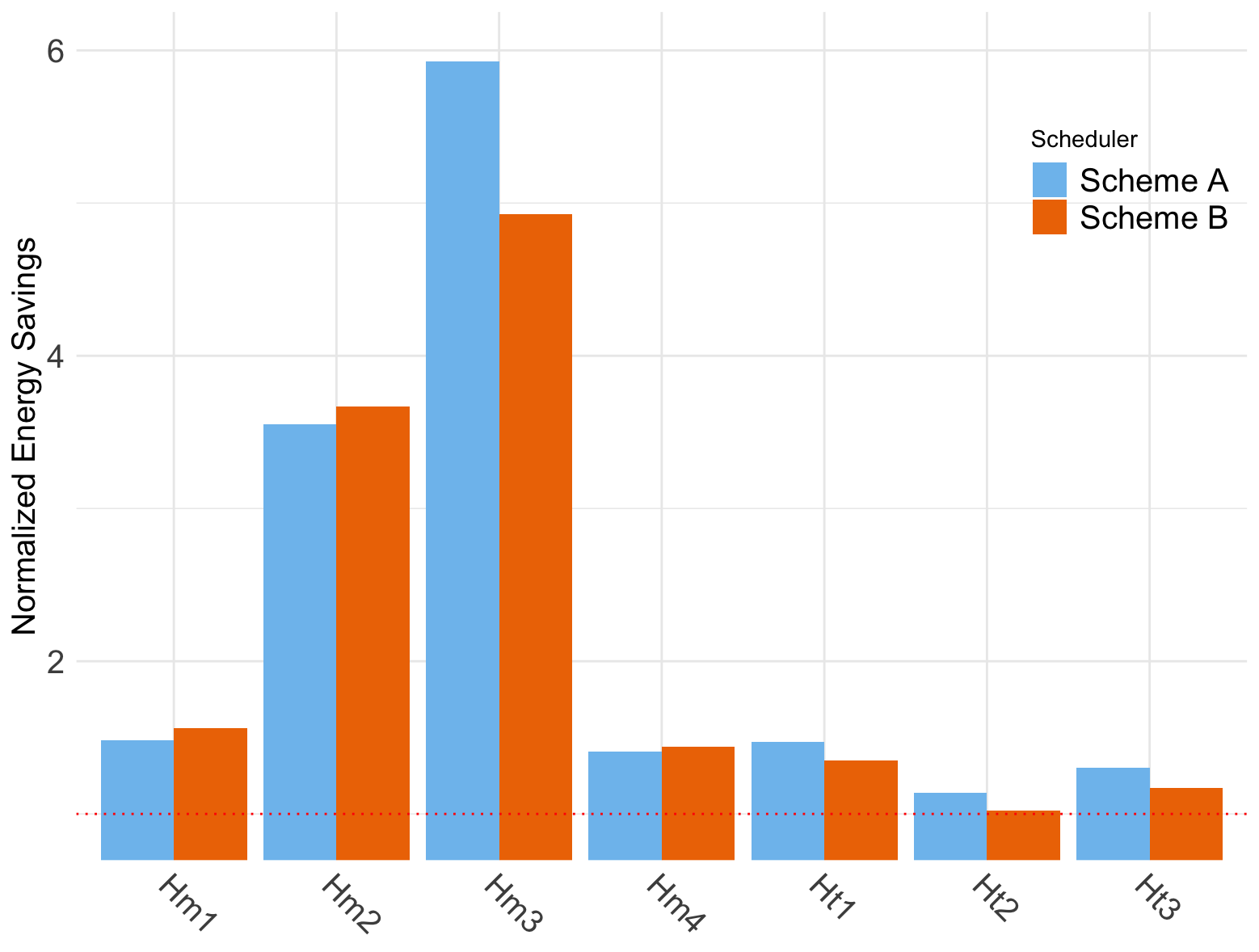}
      \caption{Energy savings - Rodinia}
      \label{figure:energy-rodinia}
  \end{subfigure}
  \qquad
  \begin{subfigure}[b]{0.3\textwidth}
      \centering
      \includegraphics[width=\textwidth]{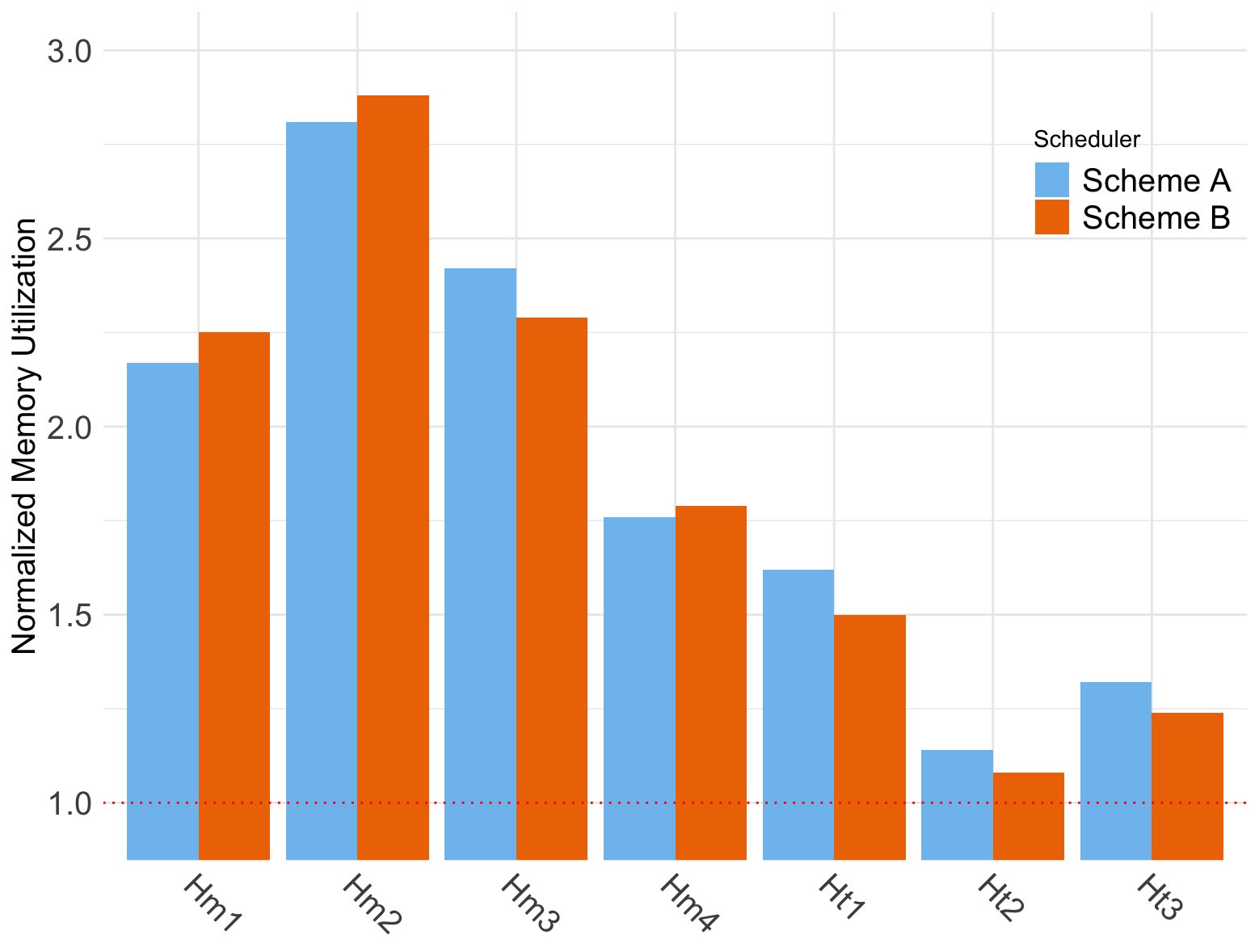}
      \caption{Memory utilization - Rodinia}
      \label{figure:memory-rodinia}
  \end{subfigure}
  \qquad
  \begin{subfigure}[b]{0.3\textwidth}
      \centering
      \includegraphics[width=\textwidth]{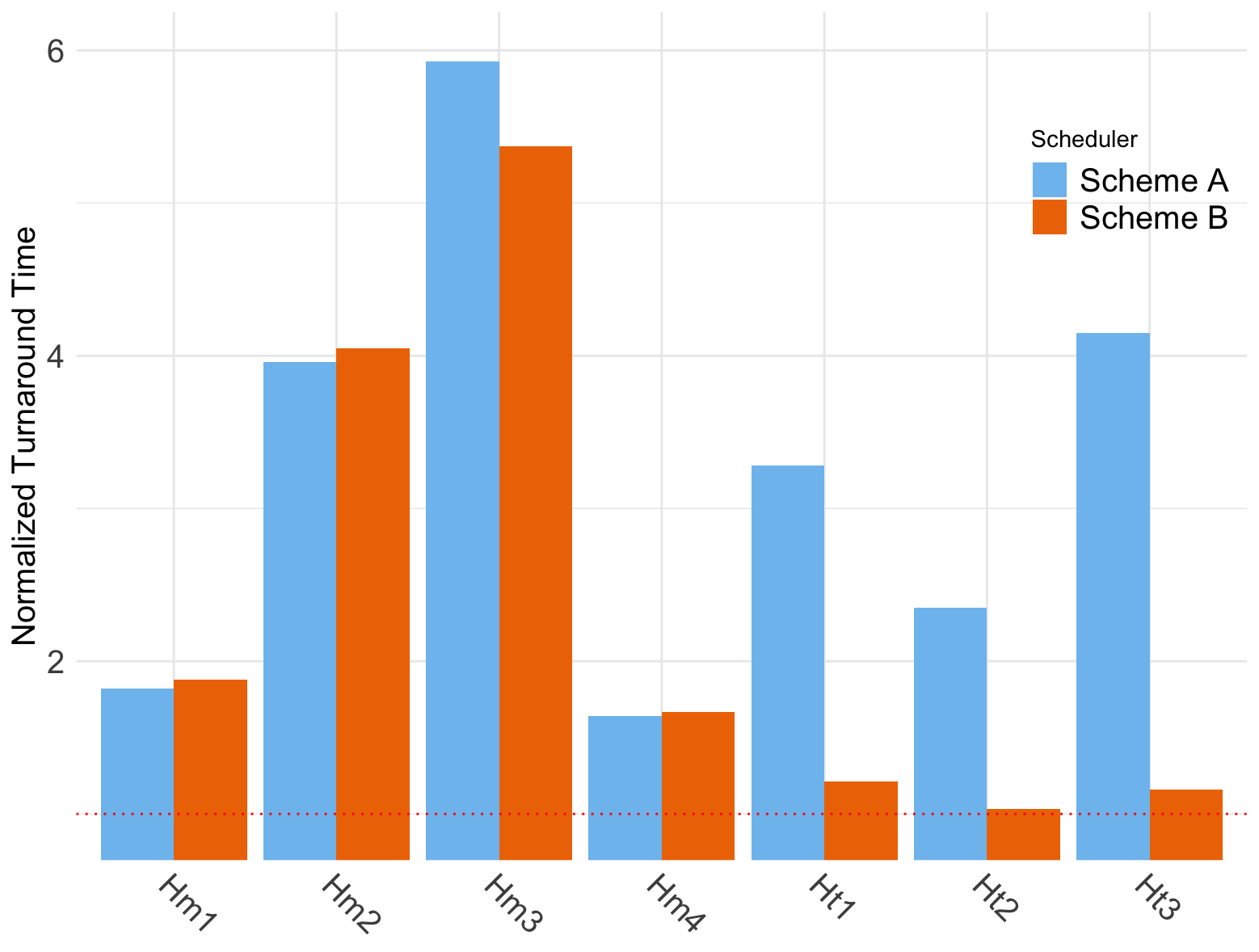}
      \caption{Job turnaround time - Rodinia}
      \label{figure:turnaround-rodinia}
  \end{subfigure}
  \qquad
  \begin{subfigure}[b]{0.3\textwidth}
  \centering
      \includegraphics[width=\textwidth]{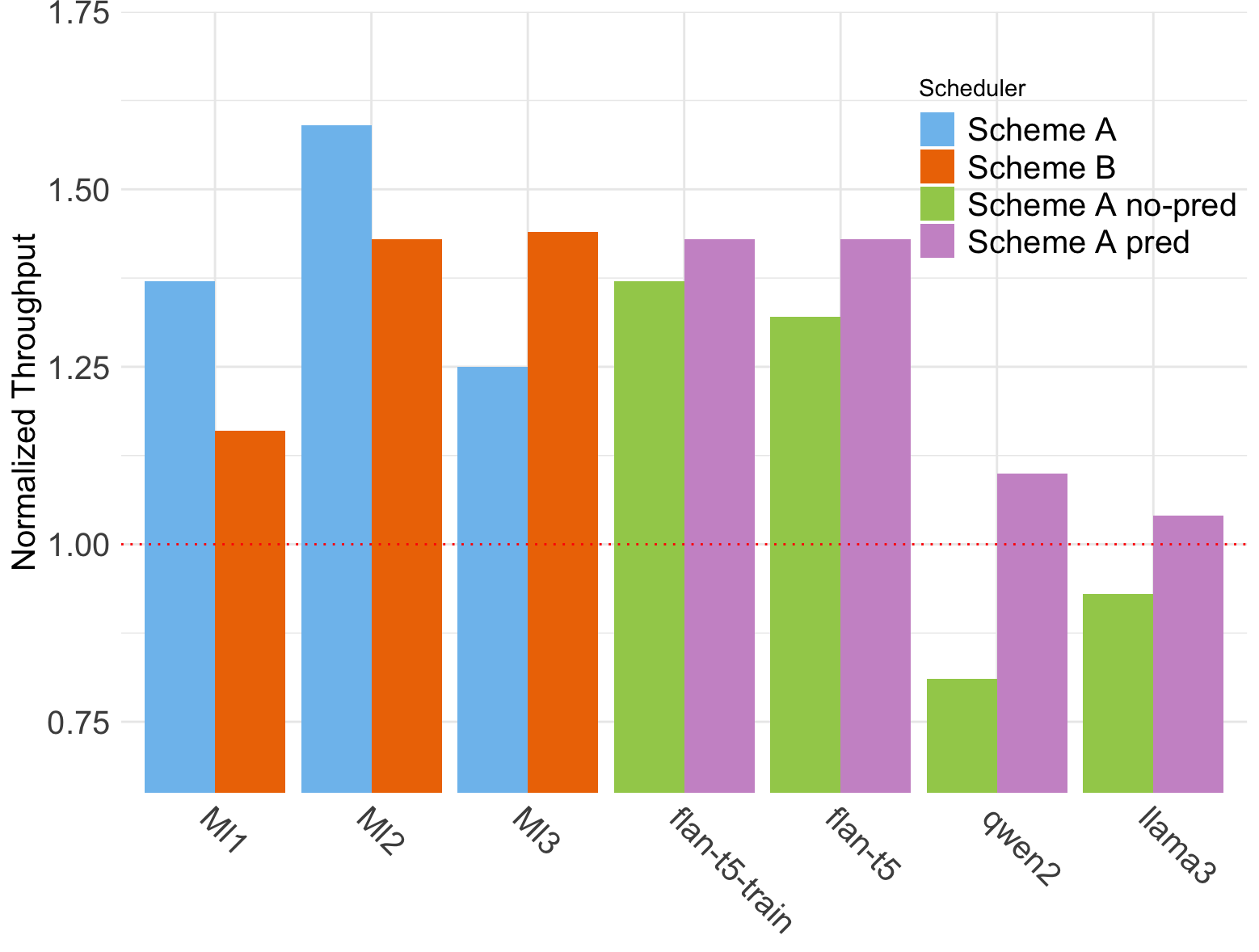}
    \caption{Throughput - ML workloads}
    \label{figure:throughput-ml}
  \end{subfigure}
  \qquad
  \begin{subfigure}[b]{0.3\textwidth}
      \centering
      \includegraphics[width=\textwidth]{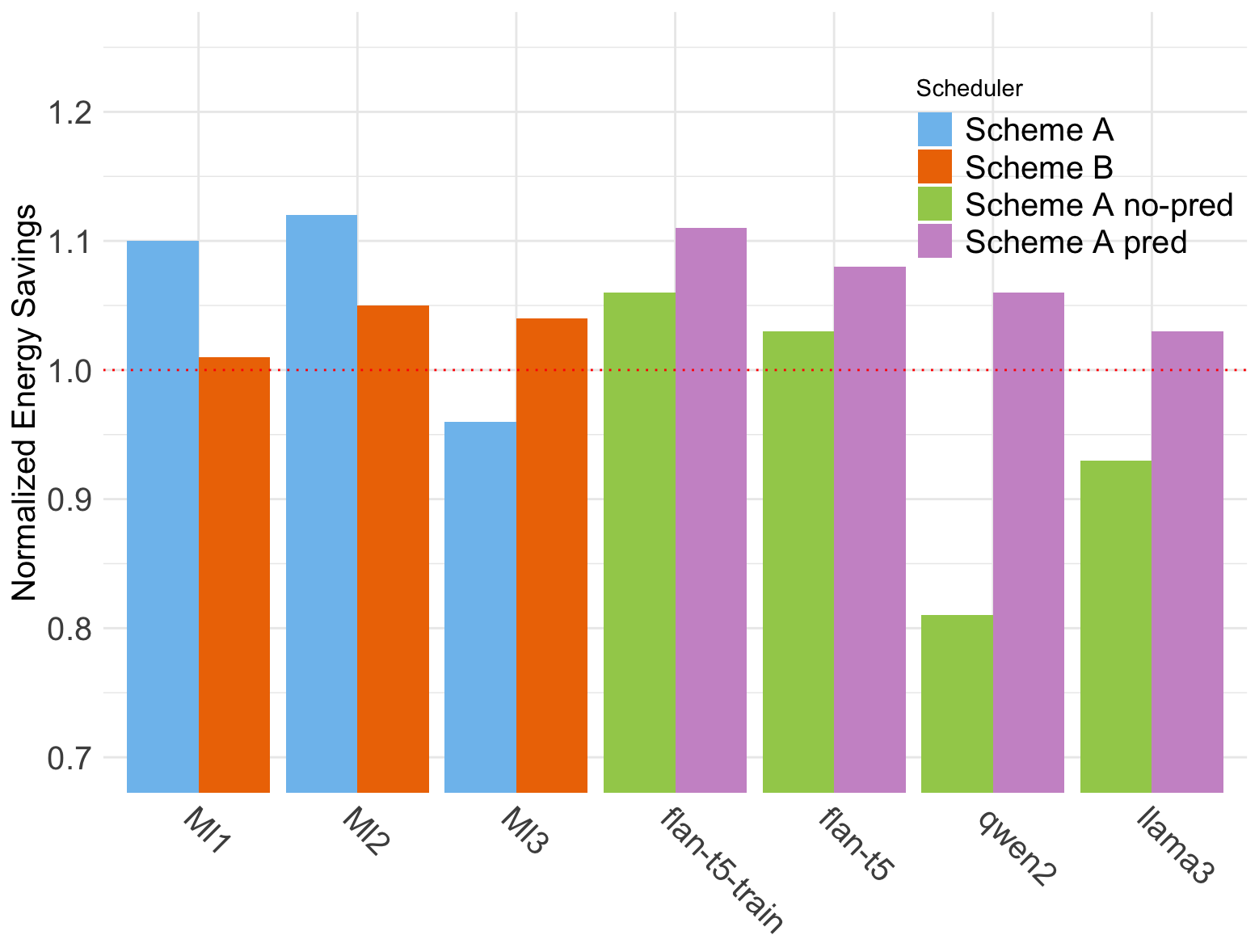}
      \caption{Energy savings - ML workloads}
      \label{figure:energy-ml}
  \end{subfigure}
  \qquad
  \begin{subfigure}[b]{0.3\textwidth}
      \centering
      \includegraphics[width=\textwidth]{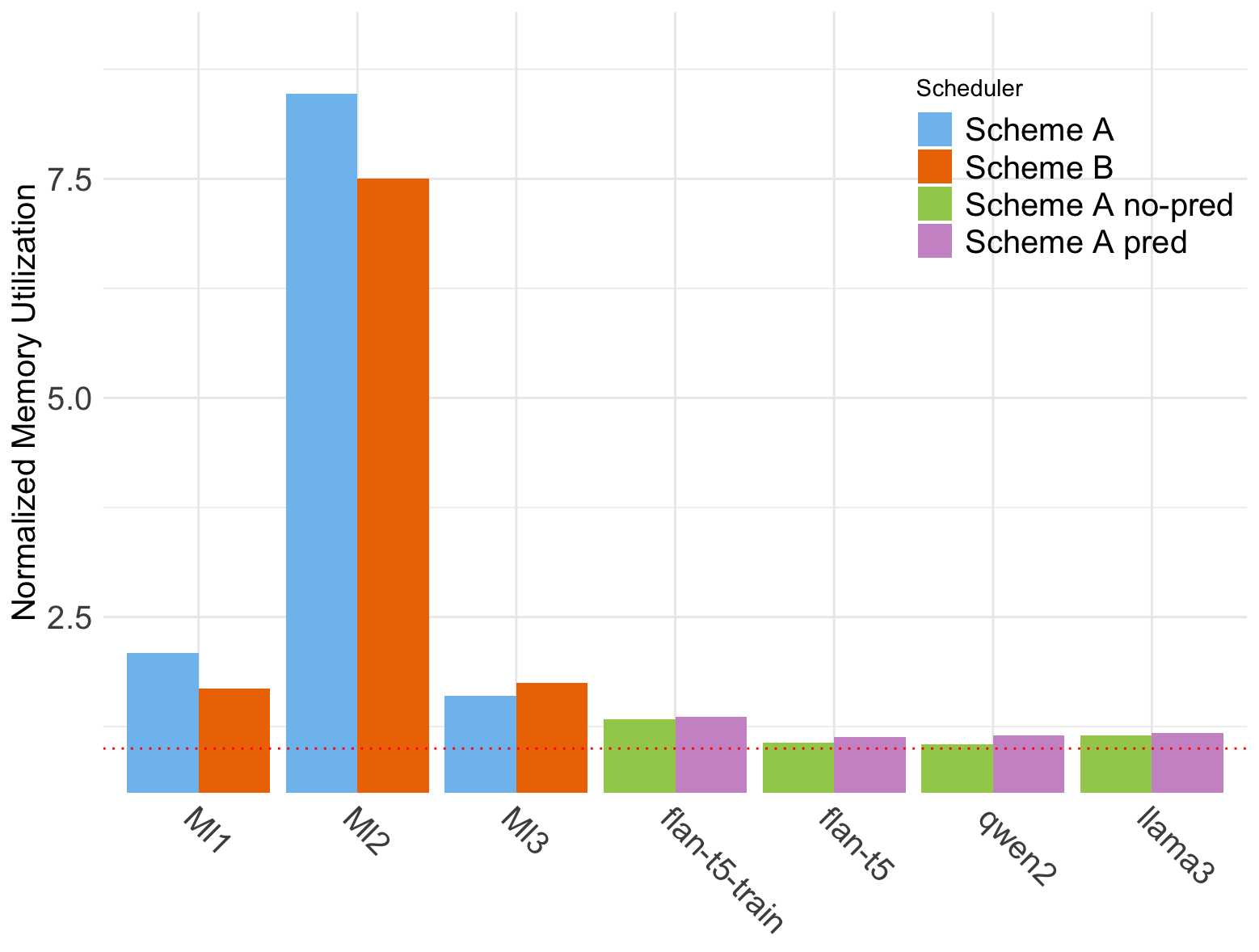}
      \caption{Memory utilization - ML workloads}
      \label{figure:memory-ml}
  \end{subfigure}
  \qquad
  \begin{subfigure}[b]{0.3\textwidth}
      \centering
      \includegraphics[width=\textwidth]{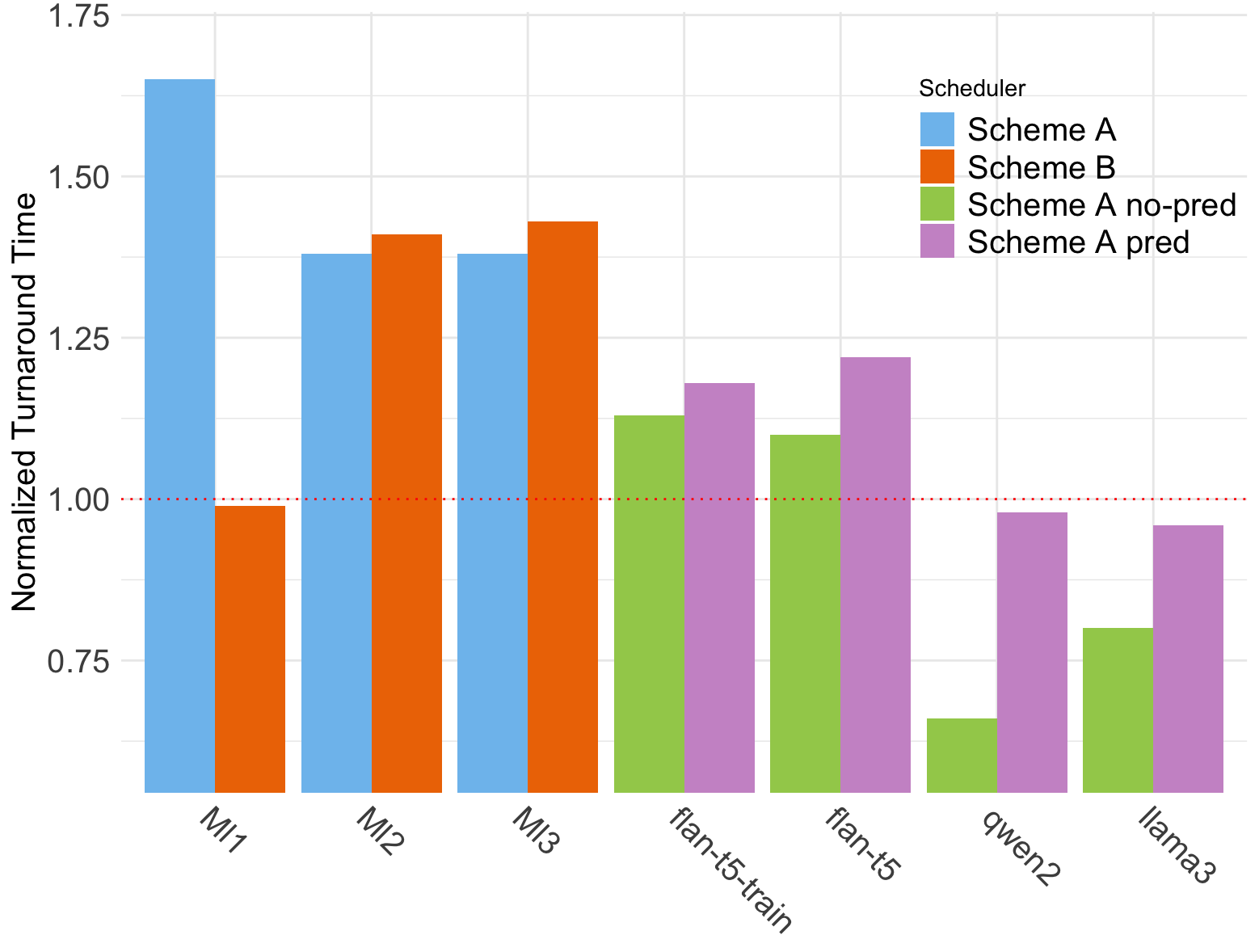}
      \caption{Job turnaround time - ML workloads}
      \label{figure:turnaround-ml}
  \end{subfigure}
  \vspace{-.1in}
  \caption{Normalized performance results on Rodinia and ML workloads.}
  \label{figure:results}
\end{figure*}

Prediction improves throughput across all workloads primarily by supporting a grow-on-demand strategy. Every job is initially placed in the smallest partition to maximize parallelism. The prediction mechanism then detects whether this allocation will be insufficient and triggers an early resize before the job encounters an OOM error. This prevents wasted runtime while still keeping the initial packing density high. As shown in Figure~\ref{figure:throughput-ml}, dynamic workloads achieve an average throughput improvement of 25.13\% compared to the baseline.
For energy saving, by preventing OOM restarts and reducing idle time through efficient partition use, prediction lowers energy per job by 6.96\% on average, as shown in Figure~\ref{figure:energy-ml}.
As shown in Figure~\ref{figure:memory-ml}, prediction achieves an average utilization improvement of 20.73\% across dynamic workloads. This benefit comes from starting jobs in the smallest partition and resizing only when necessary, which keeps GPU memory more closely matched to total actual demand.



A key advantage of dynamic memory prediction is that it intervenes before jobs actually encounter an OOM error, thereby saving substantial running time. As shown in Figures~\ref{figure:throughput-ml}–\ref{figure:turnaround-ml}, Policy A with prediction consistently outperforms Policy A without prediction. For example, in the Qwen2 benchmark, the predictor estimates that peak memory usage will exceed 10 GB as early as batch 6, whereas the job without prediction would only fail due to OOM at batch 94. For Llama-3 model, we can predicts the OOM error at batch 6 istead of hitting in at batch 72. Similarily, for flan\_t5 traning benchmark, we can predict the OOM on batch 31 instead of hitting the real OOM error in batch 41. For inference, we can predict at 21 instead of hitting the OOM at batch 27. By resizing proactively, prediction avoids nearly the entire wasted execution span, resulting in significant efficiency gains.

To evaluate the quality of the predictor, we compare its estimate at 10\% of the total iterations with the actual observed peak memory. Across workloads, the average prediction error is 14.98\% across 4 dynamic workload benchmarks. For example, in the Qwen2 benchmark, the predictor forecasts a peak 11.41GB memory usage and the final peak memory usage is 12.23GB. For Llama-3 the peak prediction is 16.64GB and final peak usage is 16.63GB. The stability of these early predictions demonstrates that the predictor not only reacts quickly but also provides results that closely track true memory demand.

\section{Related Work}\label{sec-related-work}


The MISO framework \cite{miso} aims to boostGPU utilization leveraging MIG capabilities. When a new job is slated to start, it must first run a portion
of its execution on a MIG slice with another job, which breaks
isolation guarantees (security and performance). This defeats the purpose of MIG - in real world it is of paramount important to maintain security and performance isolation of every worklooad.
Secondly, it has no automatic mechanism for managing
out-of-memory errors; a user must specify a minimum size to avoid this.
When repartitioning the GPU, it checkpoints
all active jobs and then restores them on the appropriate slices. This can be a significant overhead due to the volume of data involved.
In contrast, \shortname does not require checkpointing (opting
instead for quick restarts); does not require "training executions"; never co-locates jobs on the same MIG
slice; leverages prediction for memory footprint estimation and early restarts rather
than predicting the speedup of a job on each possible MIG slice; and
employs clever partition management to maximize concurrency (rather than seeking an
optimal configuration each time a new job arrives).

\cite{clover} attempts to reduce power consumption and carbon emissions
when hosting computationally intensive ML inference models.
It is designed specifically for an inference runtime system where
a mixture of models with varying accuracy quality are available
for serving inference requests. It leverages MIG reconfiguration
to balance the tradeoffs among carbon emissions, inference accuracy,
and SLA targets.

In \cite{gpu-pool} the authors use prediction to co-locate jobs
without degrading quality of service,
but this is aimed at full GPUs (not MIG partitions), so there is no
consideration of dynamic reconfiguration.
Another closely related work is \cite{serving-dnn-mig}, which defines
the dynamic reconfiguration scheduling problem for MIG as a
``reconfigurable machine scheduling'' (RMS) problem.
Two key differences are that it designed to work for DNNs and for Kubernetes and they are not comparable in terms of techniques developed here.

Prior to MIG support, multiple lines of research explored
GPU sharing, as it has long been a critical problem area.
These include OS-level approaches \cite{KMMB2012,RCSR2011};
techniques for preemption on the GPU
via kernel slicing
\cite{BaKa2012,SaWB2013,TGCR2014,PaPM2015,ZhTL2015,flep};
or better packing schemes, for example \cite{case}.
DNN-specific approaches also abound, e.g. \cite{zico},
which exploits the cyclic nature of DNN training's forward
and backward passes to achieve more effective job co-location and
memory usage; it increases performance by overlapping forward
passes (memory intensive) with backward passes (less memory intensive)
in hyperparameter tuning; it focuses solely on DNN training and not on utilizing MIG for generic workloads. 
Unlike the above approaches, MIGPRO focuses first on the hard problem
of memory estimation and uses this information for dynamic memory
ML workloads. \shortname's scheduler uses this information for
early predictions and to avoid late restarts. The partition manager
cleverly manages the MIG slices, performing partition fusion and
fission to create the tightest partitions. The scheduling scheme
avoids costly reconfigurations. The end result is significant
throughput improvement and energy saving. 

\section{Conclusion}\label{sec-conclusion}
In this work, we propose a comprehensive framework
called \shortname to effectively share MIG devices.
\shortname focuses first on the hard problem of
memory estimation, incorporating compiler analysis,
model size estimation, and a time series-based
predictor for different types of workloads.
The scheduler and partition manager then use this information
to cleverly manage the MIG device, performing fusion and fission
operations to create tight partitions.
Empirical results show improvements to
throughput, energy consumption, memory utilization, and
job turnaround time.
Due to its dynamic memory predictive capability,
\shortname is able to handle modern LLM workloads. Such
attributes make \shortname an attractive candidate for scheduling generic workloads from different domains using a unified system in an effective manner to boost throughput, save energy and increase GPU utilization. 

\newpage

\bibliographystyle{plain}
\bibliography{refs.bib}

\appendix
\section{Appendix}\label{sec:appendix}
\subsection{Workload Details}\label{sec:workload-details}
There are 7 Rodinia mixes, as shown in Table \ref{tab:rod-mixes}.
The first four rows represent homogeneous mixes; the last three
rows are heterogeneous mixes, and the ratio of small:medium:large
is shown. Ht1 is an exception, as its jobs are intentionally designed
so that the small jobs together have an equal runtime to that of
the medium jobs, and similarly to that of the large jobs. The mix
in this case is 15 total, with 11 small, 2 medium, and 2 large jobs.
The other jobs in the heterogeneous mixes are chosen randomly
from a pool of Rodinia benchmark+parameter pairs.
The small jobs' memory footprints
fit within 5GB; medium within 20GB; and large within the full 40GB
of an A100.

\begin{table}[h]
    \centering
    \caption{The Rodinia job mixes used in the experiments.}
    \begin{tabular}{|c|c|c|c|}
    \hline
         \bf Mix & \bf Type & \bf Jobs & \bf Batch Size \\\hline
         Hm1 & Homogeneous  & particle filter & 50 \\\hline
         Hm2 & Homogeneous  & gaussian & 50 \\\hline
         Hm3 & Homogeneous  & myocyte & 100 \\\hline
         Hm4 & Homogeneous & euler3D & 50 \\\hline
         Ht1 & Heterogeneous & -:-:- & 15 \\\hline
         Ht2 & Heterogeneous & 1:0:1:1 & 18 \\\hline
         Ht3 & Heterogeneous & 4:0:1:1 & 36 \\\hline
    \end{tabular}
    \label{tab:rod-mixes}
\end{table}

We run 7 ML workload mixes, as shown in Table \ref{tab:ml-mixes}.
The first three rows represent mixes created randomly
from the computer vision and natural language procesing
models VGG16, ResNet50, InceptionV3, and BERT.
These are training workloads.
The last 4 rows represent homogeneous workloads of an
LLM model: FLAN-T5, Qwen 2, or Llama 3. The are
inference workloads (except in the case of FLAN-T5-train,
as indicated).

\begin{table}[h]
    \centering
    \caption{The ML mixes used in the experiments.}
    \begin{tabular}{|c|c|c|c|}
    \hline
         \bf Mix & \bf Type & \bf Jobs & \bf Batch Size \\\hline
         Ml1 & Heterogeneous  & 1:0:1:0 & 14 \\\hline
         Ml2 & Heterogeneous  & 1:0:0:0 & 21 \\\hline
         Ml3 & Heterogeneous  & 0:0:1:0 & 18 \\\hline
         FLAN-T5-train & Homogeneous & flan-t5 & 4 \\\hline
         FLAN-T5 & Homogeneous & flan-t5 & 6 \\\hline
         Qwen2 & Homogeneous & qwen2 & 1 \\\hline
         Llama 3 & Homogeneous & llama3 & 1 \\\hline
    \end{tabular}
    \label{tab:ml-mixes}
\end{table}

Table \ref{tab:myocyte-breakdown} records the timing of the benchmark used in the homogeneous mix Hm1. We use this to show that MIG slices incur possible overheads in memory management as each MIG slice has its own address space. 
\vspace{-1em}
\begin{table}[!h]
\centering
\caption{Myocyte Run breakdown, Scheme A (1/7 Compute, 1/8 Memory) vs. Baseline (Full GPU)}
\resizebox{\columnwidth}{!}{%
\begin{tabular}{|c|c|c|}
    \hline
    \bf Metric & \bf Scheme A (7x1g.5gb slice) & \bf Baseline (Full GPU) \\\hline
    Allocate CPU/GPU Mem  & 0.98 s & 0.24 s \\\hline
    Read data and copy to GPU Mem  & 0.0102 s & 0.0122 s \\\hline
    GPU kernel runtime  & 0.002647 s & 0.003555 s \\\hline
    Copy data from GPU to CPU & 3.47 s & 3.36 s \\\hline
    Free GPU Memory & 0.02469 s & 0.00058 s \\\hline
\end{tabular}%
}
\label{tab:myocyte-breakdown}
\end{table}

\begin{table}[!h]
    \centering
    \caption{Needleman-Wunsch, Baseline (Full GPU) vs Policy A 7x(1/7 Compute, 1/8 Memory)}
    \resizebox{\columnwidth}{!}{%
    \begin{tabular}{|c|c|c|}
    \hline
         \bf Metric & \bf Policy A (7x1g.5gb slice) & \bf Baseline (Full GPU) \\\hline
        Single Benchmark Runtime (microseconds) & 1171507 & 523406 \\\hline
    \end{tabular}%
    }
    \label{tab:needle}
\end{table}


\end{document}